\documentstyle[12pt]{article}

\newcommand{\bmat}{\left(\begin{array}}
\newcommand{\emat}{\end{array}\right)}
\def\NPB#1#2#3{Nucl. Phys. B{#1} (19#2) #3}
\def\PLB#1#2#3{Phys. Lett. B{#1} (19#2) #3}

\def\PRD#1#2#3{Phys. Rev. D{#1} (19#2) #3}

\def\MODA#1#2#3{Mod. Phys. Lett.  {#1} (19#2) #3}

\def\yzero{\smash{\hbox{$y\kern-4pt\raise1pt\hbox{${}^\circ$}$}}}
\def\ov{\overline}
\def\s2{\frac{1}{\sqrt2}}

\def\beq{\begin{equation}}
\def\eeq{\end{equation}}
\def\beqa{\begin{eqnarray}}
\def\eeqa{\end{eqnarray}}

\def\Tr{{\rm Tr \,}}
\def\diag{{\rm diag \,}}

\def\Dsl{\,\raise.15ex\hbox{/}\mkern-13.5mu D} 

\def\IC{\bf C}
\def\IZ{\bf Z}
\def\IP{\bf P}

\def\IT{{\bf T}}
\def\NN{{\cal N}}

\def\id{{\bf 1}}
\def\Dfive{${\ov{\rm D5}}$}
\def\Dseven{${\ov{\rm D7}}$}
\def\Dtseven{${\widehat{\rm D7}}$}

\newcommand{\drawsquare}[2]{\hbox{%
\rule{#2pt}{#1pt}\hskip-#2pt
\rule{#1pt}{#2pt}\hskip-#1pt
\rule[#1pt]{#1pt}{#2pt}}\rule[#1pt]{#2pt}{#2pt}\hskip-#2pt
\rule{#2pt}{#1pt}}

\newcommand{\fund}{\raisebox{-.5pt}{\drawsquare{6.5}{0.4}}}
\newcommand{\Ysymm}{\raisebox{-.5pt}{\drawsquare{6.5}{0.4}}\hskip-0.4pt%
        \raisebox{-.5pt}{\drawsquare{6.5}{0.4}}}
\newcommand{\Yasymm}{\raisebox{-3.5pt}{\drawsquare{6.5}{0.4}}\hskip-6.9pt%
        \raisebox{3pt}{\drawsquare{6.5}{0.4}}}
\newcommand{\antifund}{\overline{\fund}}
\newcommand{\bYasymm}{\overline{\Yasymm}}
\newcommand{\bYsymm}{\overline{\Ysymm}}

\topmargin
-1.5cm
\textwidth
15.5cm
\textheight
24cm
\oddsidemargin
0.7cm
\evensidemargin
1.2cm

\begin{document}

\makeatletter
\@addtoreset{equation}{section}
\makeatother
\renewcommand{\theequation}{\thesection.\arabic{equation}}
\pagestyle{empty}
\rightline{CERN-TH/2000-331}

\rightline{\tt hep-th/0011048}
\vspace{2.5cm}
\begin{center}
\Large{\bf D-brane probes, RR tadpole cancellation \\
and K-theory charge\\[10mm]}
{\normalsize
Angel~M.~Uranga \footnote{\tt angel.uranga@cern.ch}\\[2mm]
{\em Theory Division, CERN}\\
{\em CH-1211 Geneva 23, Switzerland} \\[4mm]}

\vspace*{1.5cm}

\small{\bf Abstract} \\[7mm]
\end{center}

\begin{center} \begin{minipage}[h]{14.0cm}
{\small
We study RR charge cancellation consistency conditions in string
compactifications with open string sectors, by introducing D-brane probes
in the configuration. We show that uncancelled charges manifest as chiral
gauge anomalies in the world-volume of suitable probes. RR tadpole
cancellation can therefore be described as the consistency of the
effective compactified theory not just in the vacuum, but also in all
topological sectors (presence of D-brane probes). The result explains
why tadpole cancellation is usually much stronger than anomaly cancellation of 
the compactified theory (in the vacuum sector). We use the probe criterion
to construct consistent six-dimensional orientifolds of curved K3 spaces,
where usual CFT techniques to compute tadpoles are not valid. As a last
application, we consider compactifications where standard RR charge
cancels but full K-theory charge does not. We show the inconsistency of
such models manifests as a global gauge anomaly on the world-volume of
suitable probes.
}

\end{minipage}
\end{center}

\bigskip

\bigskip

\leftline{CERN-TH/2000-331}

\leftline{November 2000}

\newpage
\setcounter{page}{1}
\pagestyle{plain}
\renewcommand{\thefootnote}{\arabic{footnote}}
\setcounter{footnote}{0}

\section{Introduction}

A complete set of consistency conditions for superstring theories with
open string sectors seems to be provided by the requirements of open-closed
world-sheet duality, and cancellation of RR tadpoles. The latter condition
admits several interpretations from diverse points of view. From the
wolrd-sheet perspective, studied in \cite{caipol}, uncancelled RR tadpoles
lead to inconsistency due to world-sheet superconformal anomalies. From
the ten-dimensional spacetime viewpoint, cancellation of tadpoles corresponds 
to consistency of the equations of motion for certain (unphysical) RR
fields, or equivalently to cancellation of RR charge under $p$-form fields
with compact support. An alternative spacetime interpretation is
that RR tadpole cancellation ensures cancellation of chiral anomalies in
the low-energy effective theory. The connection between tadpoles and
anomalies first arose in the context of ten-dimensional type I theory
\cite{gs}, and is confirmed by the construction of anomaly-free type IIB
orientifolds \cite{sagncargese,dlp} in lower dimensions (see
\cite{prasagn,horava,bisagn} for early references, and 
\cite{gp,gjdp,fourdim,afiv} for recent constructions). More direct 
comparisons in compactified models have been performed in \cite{abiu,
morales}.

In this paper we introduce D-brane probes in diverse compactifications
with open string sectors, and explore aspects of the above two spacetime
interpretations of RR charge cancellation conditions. In the first place,
in compactifications below ten dimensions, RR tadpole cancellation
conditions are generally much stronger than anomaly cancellation in the
compactified effective theory. This is obvious in compactifications with
non-chiral spectrum, but also holds for many chiral compactifications
\cite{abiu} (see also \cite{morales}). We show in explicit examples that
uncancelled RR tadpoles manifest as gauge anomalies on the world-volume of
suitable D-brane probes. Interpreting the probes as topological defects in
the effective compactified theory, one can argue that full cancellation of
RR tadpoles corresponds to consistency of the compactified effective
theory not just in its vacuum sector, but in all possible topological
sectors. We find this viewpoint interesting, in that it can be applied in
quite general situations (curved internal spaces, non-perturbative vacua)
where other techniques are not valid. In fact we employ the probe
criterion to construct consistent six-dimensional orientifolds of curved
K3 manifolds. Also, it emphasizes the viewpoint of the compactified
theory, and hence it can be applied to theories with non-geometric
internal CFT's, like the asymmetric orientifolds in \cite{asym}.

A second direction we explore is the relevance of the fact that D-brane
charge is actually classified by K-theory \cite{mm,witten}, which differs
from the naive classification by cohomology in torsion pieces, i.e. the
existence of additional discrete charges, typically associated to stable
non-BPS branes (see \cite{revnbps} for a review). Detailed analysis of the
K-theoretical nature of RR fields \cite{mw} imply that consistency of a
model requires cancellation of the full K-theory charge, and not just of
the naive part. In fact, we consider some explicit compactifications where
standard RR charge cancels but the full K-theory charge does not, and
study them using D-brane probes. We show that the inconsistency of the
model manifests as global gauge anomalies on the world-volume of suitable
probes. Therefore D-brane probes provide a simple technique to verify
cancellation of K-theory charge in complicated models, where direct
analysis would be untractable. For instance, using suitable probes we
derive certain K-theory consistency conditions (discussed in
\cite{blpssw} from a different viewpoint) in the type IIB $\IT^4/\IZ_2$ 
orientifold in \cite{bisagn,gp}.

The paper is organized in several Sections which can be read independently, 
and we advice readers interested in just one topic to safely skip unrelated
discussions. In Section~2 we present some warm-up examples of the
introduction of D-brane probes in simple open string compactifications,
and verify the relation between RR tadpole cancellation and consistency of
the probe world-volume theories. In Section~3 we apply this technique to
study six- and four-dimensional toroidal $\IT^4/\IZ_N$
and $\IT^6/\IZ_N$ type IIB orientifolds. We show that cancellation of RR
tadpoles unrelated to anomaly cancellation in the compactified theory can
however be recovered by requiring consistency of suitable D-brane probes.
Hence, tadpole conditions are equivalent to consistency of the compactified 
effective theory in all possible topological sectors. In Section~4 we
construct certain six-dimensional orientifolds of type IIB theory on
curved K3 spaces, and use the probe criterion to check cancellation of RR
charges in the model. In Section~5 we use D-brane probes to study
compactifications where naive RR tadpole cancellation is satisfied, but
full K-theory charge does not cancel. We show that such theories are
inconsistent, and that their inconsistency manifests as a global gauge
anomaly on suitable D-brane probes. We also find that the `non-perturbative' 
consistency conditions proposed in \cite{blpssw} for the model in
\cite{bisagn,gp} are of K-theoretic nature, and rederive them using
D-brane probes. Section~6 contains some final remarks.

\section{Warm-up examples}

\subsection{Toroidal compactification of type I}

The simplest example of open string theory compactification where RR tadpole 
cancellation is not related to anomaly cancellation in the lower-dimensional 
theory, is toroidal compactifications of type I theory. Hence, we consider
type IIB on $\IT^m$ modded out by world-sheet parity $\Omega$, and introduce 
a number $N$ of background D9-branes, which is constrained to be $N=32$ by
RR tadpole cancellation. For future use, we have in mind $m=4$ or $m=6$,
even though the argument applies more generally. One obtains a
$(10-m)$-dimensional non-chiral theory, with sixteen supersymmetries and a
rank 16 gauge group, equal to $SO(32)$ if no Wilson lines are turned on.
From the compactified effective theory viewpoint, there seems to be no
natural explanation for this specific rank, i.e. for the requirement of
having 32 background D-branes, and the compactified theory would seem to
make sense even for other choices.

However, this constraint can actually be understood even from the compactified 
theory viewpoint, by considering it in topologically non-trivial sectors,
like in the background of a string-like topological defect with charge
$n$, realized in string theory as a stack of $n$ D1-brane probes spanning
two of the non-compact dimensions
\footnote{Notice that the probes do not introduce new RR tadpoles, or
rather their flux is allowed to escape to infinity along transverse
non-compact dimensions, hence $n$ is unconstrained. Also, for $m=6$, the
D1-branes induce a deficit angle and induce asymptotic curvature. To avoid
this as an objection, one may introduce as a probe a set of well separated
D1-branes and anti-D1-branes. Despite lack of supersymmetry and stability,
they form a consistent probe, for which world-volume anomalies should
cancel. Their analysis is similar to that with only D1-brane probes, hence
we phrase the discussion in the latter terms.}. We may use string theory
techniques to compute the zero modes of this soliton, which is a simple 
generalization of that in \cite{polwi}. The two-dimensional world-volume
field theory has $(0,8)$ supersymmetry (see \cite{twodim,gcu} for the
multiplet structure of two-dimensional theories with diverse
supersymmetries). In the $11$ sector we obtain a $(0,8)$ $SO(n)$ gauge
multiplet (containing gauge bosons and eight left-handed Majorana-Weyl
(MW) fermions), and one $(0,8)$ chiral multiplet (with eight real scalars
and eight right-handed MW fermions) in the two-index symmetric 
representation. The $19+91$ sectors contain (regardless of the possible
existence of D9-brane Wilson lines) one left-handed MW fermion in the
$(n,N)$ bi-fundamental representation (Fermi multiplet). 

Cancellation of gauge anomalies requires $N=32$, and therefore shows the
necessity of imposing the RR tadpole cancellation condition in order to
obtain a consistent probe. Equivalently, in order for our compactified
theory to be consistent in the relevant topological sector.

Of course, a simpler argument, suggested in \cite{morales}, is to notice
that the full theory contains gauge and gravitational degrees of freedom
propagating in ten dimensions, and their ten-dimensional anomalies should
cancel \footnote{From the lower-dimensional viewpoint, this corresponds to
the existence of certain Ward identities for amplitude involving KK
excited states.}, hence $N=32$. As explained in the introduction, we do
not claim the probe argument is the only reason why string theory requires 
RR tadpole cancellation. However, we find the probe approach is an
interesting explanation, which may be easier to apply in certain
situations, and which provides an understanding of tadpole conditions
from the lower-dimensional viewpoint.

\smallskip

A similarly simple example is type II theory compactified on $\IT^m$ modded
out by $\Omega R$, where $R$ reverses all internal coordinates (and contains 
$(-1)^{F_L}$ if required). We take IIB theory for $m$ even and IIA
for $m$ odd. These models contain a number $N$ of D$(9-m)$-branes, with
$N=32$ in order to cancel RR tadpoles. These theories are related to
toroidal compactifications of type I by T-duality along the internal
dimensions. Therefore, from the point of view of the compactified theory,
the requirement $N=32$ can be detected by considering a string-like defect
(which we denote `fat string') represented in string theory as a stack of
$n$ D$(m+1)$-brane wrapped on $\IT^m$. The computation of the zero modes
on the fat string is isomorphic to that above, and $N=32$ is again obtained 
as the condition of cancellation of two-dimensional gauge anomalies.

\smallskip

Although simple, these results will be quite helpful in the study of more
involved orientifold models, like six- or four-dimensional orientifolds of 
$\IT^4/\IZ_N$ or $\IT^6/\IZ_N$, whose discussion we postpone until Section
3.

\subsection{Configurations of D6-branes on a general Calabi-Yau}

It is easy to extend the probe argument to more general compactifications
containing D-branes. For instance, let us consider type IIA theory
compactified on a Calabi-Yau threefold ${\bf X}_3$, with a set of
D6-branes
labeled by an index $a$, in stacks of multiplicity $N_a$ and wrapping a
set of 3-cycles $\Pi_a$. We take these 3-cycles to be special lagrangian,
and hence supersymmetric \cite{bbs}. Antibranes are treated on an equal
footing with branes by simply considering the wrapped cycle with the
opposite orientation. Configurations of D6-branes on 3-cycles on Calabi-Yau 
threefolds have appeared in a number of references \cite{vafakarch,kachru}. 
The particular case of the six-torus appears in \cite{afiru}. For simplicity 
we do not consider the introduction of orientifold actions (see \cite{bgkl} 
for the toroidal case).

Since the D6-branes are sources for the RR 7-form Hodge dual to the IIA
1-form, with charge proportional to the homology class $[\Pi_a]$ of the
wrapped cycle, the tadpole cancellation condition is the vanishing of the net
charge in the compact space,
\beqa
\sum_a N_a [\Pi_a] = 0
\label{homology}
\eeqa
From the viewpoint of the compactified effective theory, these conditions do
ensure the cancellation of four-dimensional anomalies, but are in fact much
stronger conditions. The field theory in four dimensions has the following
structure. The $6_a6_a$ sector contains fields in multiplets of the $\NN=1$ 
supersymmetry preserved by the D6$_a$-brane. It produces $U(N_a)$ vector
multiplets, and a set of additional chiral multiplets in the adjoint 
representation if the 3-cycle $[\Pi_a]$ is not rigid, which do not generate 
chiral anomalies. In the mixed $6_a6_b$ and $6_b6_a$ sectors we obtain
chiral left-handed fermions in the representation $\sum_{a,b} I_{ab}
(N_a,{\ov N}_b)$, where $I_{ab}=[\Pi_a]\cdot[\Pi_b]$ is the intersection 
number, and a negative multiplicity corresponds to a positive multiplicity
of opposite chirality fields (see \cite{bgkl,afiru} for a discussion
in the toroidal context). The intersection may produce additional light 
(massless or tachyonic) scalars and vector-like fermions, depending on the
local geometry of the D6-branes and ${\bf X}_3$ around the intersection,
but these fields do not contribute to chiral anomalies.

The conditions of cancellation of cubic non-abelian anomalies \footnote{As
follows from \cite{inflow}, the local anomaly at each intersection is
cancelled by the inflow mechanism, namely the violation of charge due to
the chiral fermion anomaly at each intersection is cancelled by a charge
inflow from the branes (see \cite{inflowcomp} for string computations of
the relevant couplings). From this perspective, conditions (\ref{anomdsix})
ensure the global consistency of the inflow mechanism, namely cancellation
of inflows from different intersections into a given brane \cite{afiru}
(see \cite{hz} for a similar effect).} read
\beqa
\sum_b N_b\, [\Pi_a]\cdot [\Pi_b] = 0\;\;\; {\rm for}\;{\rm all}\; a
\label{anomdsix}
\eeqa
Since the set of cycles $[\Pi_a]$ need not be complete, this does not
necessarily imply $\sum_b N_b [\Pi_b]=0$. It simply states that the
overall homology class has zero intersection with any of the individual
homology classes. This condition can be achieved even if the total homology 
class is nonzero, leading to the construction of models where RR tadpole
conditions are not satisfied even though the inconsistency does not show
up as a breakdown of gauge invariance in the low energy field theory.

As in previous examples, it is possible to make the inconsistency manifest
even from the lower-dimensional viewpoint, by considering it in suitable
soliton backgrounds, equivalently introducing suitable probes in the
configuration. For instance, we may consider the introduction of fat
strings obtained by wrapping $n_i$ D4-branes on a 3-cycle $[\Lambda_i]$,
where $\{ [\Lambda_i]\}$ is a basis of supersymmetric 3-cycles. The field
theory on the $i^{th}$ probe has $(0,2)$ supersymmetry, and we may use
standard string theory arguments to obtain the two-dimensional world-volume 
fields. In the $4_i4_i$ sector, fields fill out $(2,2)$ multiplets, and
produce $U(n_i)$ gauge symmetries, but do not generate any two-dimensional
anomalies. In the mixed $4_i6_a$ and $6_a4_i$ sector we obtain a set of
left MW fermions in the representation $(n_i,{\ov N}_a)$, with
multiplicity $[\Lambda_i]\cdot[\Pi_a]$ (following the usual convention on
its sign). Cancellation of the two-dimensional non-abelian anomaly on the
probes leads to the constraints
\beqa
[\Lambda_i]\cdot \sum_a N_a [\Pi_a] = 0 \quad {\rm for \; all\;} i
\label{probehomology}
\eeqa
which, since $\{[\Lambda_i]\}$ is a complete set, require $\sum_a
N_a[\Pi_a]=0$, i.e. the tadpole cancellation condition (\ref{homology}).

We conclude this section with two short comments. The inclusion of
orientifold planes in the construction would lead to non-zero contributions 
on the right hand side of the tadpole equation (\ref{homology}). The
corresponding modification in (\ref{probehomology}) would arise from
chiral anomalies arising from the $4_i4_i$ sector due to the different
action of the orientifold projection of the corresponding chiral fermions.

A second comment concerns how general is the observation that RR tadpole
cancellation is equivalent to anomaly cancellation on D-brane probes. The
above models suggest the following approach, for the case of geometric
compactifications. Let $[\Pi]$ denote total RR charge arising from
orientifold planes and D-branes in the configuration (filling non-compact
spacetime completely, but possibly only partially wrapped on the internal
space), understood as an element in the homology of the internal space.
If non-zero, the inconsistency of the theory can be detected by introducing 
a fat string obtained as a D-brane with charge Hodge dual (in the internal
space) to $[\Pi]$. The non-zero intersection between classes leads to
chiral anomalies on the probe world-volume, as in the more explicit models
above. 

Recent developments indicate that D-brane charges are actually K-theory
(rather than (co)homology) classes \cite{mm,witten}. In Section~5 we
discuss the problem of charge cancellation in certain type IIB orientifold
compactification, and its interplay with D-brane probes in the configuration.

The above argument shows that in any geometric compactification RR tadpole
conditions can be equivalently described as consistency of all possible
probes of the configuration. However, many compactifications do not have a
simple (if any at all) geometric interpretation, hence it is important to
study explicitly the behaviour of probes in such cases. In next section
we center on orientifolds of type IIB toroidal orbifolds, where singularities 
in the internal space render geometrical techniques much less useful. 

\section{Six- and four-dimensional orientifolds}

\subsection{Six-dimensional $\NN=1$ orientifolds}

In this section we consider $\Omega$ orientifolds of type IIB theory
\cite{sagncargese,dlp} compactified to six-dimensions on toroidal
orbifolds $\IT^4/\IZ_N$ (see \cite{prasagn,horava,bisagn} for early work,
and \cite{gp,gjdp} for more recent references), or T-dual versions of
these models. As discussed in a number of non-compact examples
\cite{bi1,bi2}, and in \cite{abiu} in compact models, cancellation of
six-dimensional anomalies is exactly equivalent to cancellation of twisted
RR tadpoles. On the other hand, untwisted RR tadpoles are not related to
six-dimensional anomalies, even though they are required for consistency
of the theory. The necessity of the latter constraints can be made
manifest from the six-dimensional viewpoint by considering suitable
probes, as we discuss in a particular case for the sake of concreteness.

Let us consider the $\Omega$ orientifold of type IIB theory on $\IT^4/\IZ_3$, 
studied in \cite{gjdp}. The closed string spectrum contains the $D=6$, $\NN=1$ 
supergravity multiplet, the dilaton tensor multiplet, and eleven hyper- and 
nine tensor multiplets. Cancellation of untwisted RR tadpoles requires the
introduction of 32 D9-branes, and zero net number of D5-branes. For an 
arbitrary number of D9-branes, the general form of the Chan-Paton embedding of 
the $\IZ_3$ generator $\theta$ is
\beqa
\gamma_{\theta,9} = \diag(\id_{N_0},e^{2\pi i\frac 13} \id_{N_1},
e^{2\pi i\frac 23} \id_{N_1})
\eeqa
If no Wilson lines are turned on, the six-dimensional open string spectrum
contains the following $D=6$ $\NN=1$ multiplets
\beqa
{\rm Vector} & SO(N_0)\times U(N_1) \nonumber\\
{\rm Hyper} & (\fund_0, \antifund_1) + \Yasymm_1
\eeqa
Cancellation of six-dimensional gauge and gravitational anomalies requires
$N_1=N_0+8$, which agrees with the twisted RR tadpole cancellation condition 
$\Tr \gamma_{\theta,9}=-8$, but does not fix the total number of D9-branes
in the theory. Consideration of further anomalies, like $U(1)$ anomaly
cancellation by the generalized Green-Schwarz mechanism in \cite{sagnan},
does not lead to new constraints.

Based on our previous experience, however, we argue that the six-dimensional 
field theory can actually detect the inconsistency of not imposing untwisted 
RR tadpole conditions. For instance, we may introduce a stack of $n$ D1-brane 
probes sitting at a generic point in the internal space (hence we include
their $\IZ_3$ images). The probe is insensitive to the $\IZ_3$ twist and its 
spectrum of zero modes is exactly as in section 2.1. Cancellation of
world-volume anomalies requires a total number of 32 D9-branes, $N_0+2N_1=32$, 
which is the untwisted tadpole condition. Full RR tadpole cancellation
hence fixes $N_0=8$, $N_1=16$.

The untwisted tadpole condition controlling the net D5-brane charge in the
background, if D5-branes or antibranes is allowed, can be obtained by
introducing a fat string probe, constructed by wrapping a D5-brane
(denoted D5$_p$) on the internal space, with generic continuous Wilson
lines. For $\IZ_3$ the $5_p5_p$ sector is non-chiral, and the only chiral
world-volume fermions arise from mixed $55_p$, $5_p5$ sectors of string
stretched between the probe and background D5-/\Dfive-branes, if present.
Two-dimensional anomaly cancellation on the probe requires the number of
background D5- and \Dfive-branes to be equal, hence net fivebrane charge
should be zero.

In the above argument we have used six-dimensional anomaly cancellation to
recover the twisted tadpole condition, but it is easy to recover it
using a probe as well. Consider introducing a stack of $n$ D1-brane probes
sitting at the origin in $\IT^4/\IZ_3$, with the $\IZ_3$ action embedded
as
\beqa
\gamma_{\theta,1} = \diag (\id_{n_0},e^{2\pi i\frac 13}\id_{n_1},
e^{2\pi i\frac 23} \id_{n_1})
\eeqa
with $n_0+2n_1=n$. The theory on the D1-brane world-volume has $(0,4)$
supersymmetry, and contains fields in vector multiplets (gauge bosons and
four left MW fermions), chiral multiplets (four real scalars and four
right MW fermions), and fermi multiplets (four left MW fermion),  as
follows
\beqa
\begin{array}{ccc}
{\bf Sector} & {\bf (0,4)\; multiplet} & {\bf Representation} \\
{\bf 11}     & {\rm Vector} & SO(n_0)\times U(n_1) \\
             & {\rm Chiral} & \Ysymm_0 + {\rm Adj}_1 \\
             & {\rm Chiral} & (\fund_0,\fund_1) + \bYsymm_1 \\
             & {\rm Fermi}  & (\fund_0,\fund_1) + \bYasymm_1 \\
{\bf 19+91}  & {\rm MW_L\; Fermion}  & (\fund_0; N_0) + (\fund_1;N_1)
+(\antifund_1;N_1)
\end{array}
\eeqa
Cancellation of two-dimensional anomalies requires $N_0=16$, $N_1=8$, hence 
yielding both twisted and untwisted tadpole conditions. Thus, even though
in following sections we use vacuum anomaly cancellation to partially test
consistency, the corresponding constraints can also be recovered by
suitable probes.
 
The pattern we have seen in the $\IZ_3$ orientifold holds for other similar 
six-dimensional orientifolds, supersymmetric \cite{gjdp} or not \cite{ads,au,
ru}. Six-dimensional anomaly cancellation requires cancellation of twisted
tadpoles, while cancellation of untwisted tadpoles is required by
consistency of suitable string-like configurations. \footnote{Note that for 
even order orbifolds, the $5_p5_p$ sector is chiral and anomaly cancellation 
requires 32 units of net D5-brane charge (or 32 \Dfive-branes for peculiar
orientifold actions \cite{ads,au}).}. It would be interesting to study
probes in other classes of six-dimensional type IIB orientifolds \cite{blumen, 
asym, abg}. Instead of considering these extensions, we turn to the
four-dimensional case, where the pattern of anomalies vs. tadpoles is more
interesting.

\subsection{Four-dimensional Orientifolds}

Four-dimensional orientifolds of type IIB on $\IT^6/\IZ_N$ or $\IT^6/(\IZ_N 
\times \IZ_M)$ \cite{fourdim,kak33,afiv} have the new interesting feature
that they contain {\em twisted} tadpoles whose cancellation is not related
to the cancellation of anomalies in the four-dimensional effective theory
\cite{abiu}. Such twisted tadpoles are associated to orbifold twists with
fixed planes. For e.g. a set of D3-branes sitting at a point in the fixed
plane, cancellation of chiral anomalies is a local effect, which is unrelated 
to cancellation of the twisted RR tadpole, a global constraint on the
charges distributed on the plane. In T-dual versions with branes wrapped
on the fixed planes, the twisted RR tadpole condition is related to
anomaly cancellation for fields propagating in six dimensions, and is
again unrelated to four-dimensional anomalies. In this section we would
like to clarify the interpretation of these twisted RR tadpole conditions
as anomaly cancellation conditions on suitable D-brane probes. For
concreteness we center on a prototypical case, extension to other models
being straightforward. 

We consider type IIB theory on $\IT^6/(\IZ_3\times \IZ_3)$, with generators 
$\theta$ and $\omega$ corresponding to the twist eigenvalue vectors
$v=\frac 13(1,0,-1)$, and $w=\frac 13(0,1,-1)$, respectively. We mod out
this model by the orientifold action $\Omega'\equiv \Omega R_1 R_2 R_3
(-1)^{F_L}$, where $R_i:z_i\to -z_i$. This orientifold is T-dual to that
considered in \cite{kak33,afiv}. Cancellation of untwisted tadpoles
requires the introduction of 32 D3-branes filling the four non-compact
dimensions, and no net number of D7-branes. Cancellation of twisted
tadpoles at the origin requires the D3-brane Chan-Paton matrices to
satisfy
\beqa
\Tr \gamma_{\theta\omega,3} & = & -4 \nonumber \\
\Tr \gamma_{\theta,3}& = & \Tr \gamma_{\omega,3} = \Tr
\gamma_{\theta\omega^2,3}= 8 
\eeqa
The general form of the Chan-Paton matrices, before imposing the
constraints from tadpole cancellation, is
\beqa
\gamma_{\theta,3} & = & \diag (\id_{N_{00}}, \id_{N_{01}}, \id_{N_{0,2}},
\alpha \id_{N_{10}}, \alpha \id_{N_{11}}, \alpha \id_{N_{12}}, \alpha^2
\id_{N_{20}}, \alpha^2 \id_{N_{21}}, \alpha^2 \id_{N_{22}}) \nonumber \\
\gamma_{\omega,3} & = & \diag (\id_{N_{00}}, \alpha \id_{N_{01}},
\alpha^2 \id_{N_{0,2}}, \id_{N_{10}}, \alpha \id_{N_{11}}, \alpha^2
\id_{N_{12}}, \id_{N_{20}}, \alpha \id_{N_{21}}, \alpha^2 \id_{N_{22}})
\label{cpfour}
\eeqa
The orientifold projection requires $N_{ab}=N_{-a,-b}$, with subindices
understood mod 3. We also have $\gamma_{\Omega',3}=\gamma_{\Omega',3}^T$.
The unique solution to the tadpole conditions, in the case of locating all
32 D3-branes at the origin, is $N_{00}=8$, $N_{10}=N_{01}=N_{11}=4$,
$N_{12}=0$.

It is however interesting to derive these tadpole consistency conditions
by introducing suitable probes, and requiring cancellation of world-volume
gauge anomalies. Clearly, one can easily reproduce the untwisted tadpole
conditions by working in analogy with the above six-dimensional case.
Introducing fat strings constructed by wrapping D3-branes on the $i^{th}$
complex plane, with generic Wilson lines, and sitting at a generic point
in the remaining two, one recovers the condition that no net D7$_i$-brane
charge (i.e associated to D7-branes transverse to the $i^{th}$ plane) is
allowed in the model. Introducing a D7-brane wrapped on the internal
space, with generic Wilson lines, one recovers the condition that the
number of D3-branes in the model is 32. 

Partial information about the twisted tadpole cancellation conditions can
be obtained by considering four-dimensional anomaly cancellation in the
vacuum of the compactified theory. Using the general Chan-Paton embedding
(\ref{cpfour}), the four-dimensional spectrum contains the following set
$\NN=1$ supersymmetry multiplets
\beqa
{\rm Vector} \quad & SO(N_{00}) \times U(N_{10}) \times U(N_{01}) \times
U(N_{11}) \times U(N_{12}) & \nonumber \\
{\rm Chiral} \quad &
\Yasymm_{10} + (\fund_{00},\antifund_{10}) + (\fund_{01},\antifund_{11}) +
(\fund_{11},\fund_{12}) + (\antifund_{12},\antifund_{01}) & \nonumber \\
& \Yasymm_{01} + (\fund_{00},\antifund_{01}) + (\fund_{10},\antifund_{11})
+ (\fund_{11},\antifund_{12}) + (\fund_{12},\antifund_{10}) & \nonumber \\
& \bYasymm_{11} + (\fund_{00},\fund_{11}) + (\fund_{10},\fund_{01}) +
(\antifund_{01},\fund_{12}) + (\antifund_{12},\antifund_{10})
\label{specz3z3}
\eeqa
Following \cite{abiu}, cancellation of cubic non-abelian anomalies leads
to the condition
\beqa
\Tr\gamma_{\theta\omega,3} = N_{00} - N_{01} - N_{10} - N_{11} + 2 N_{12}
= -4
\label{essential}
\eeqa
which is obviously ensured by the tadpole conditions, but is a much milder
constraint. As checked in \cite{abiu}, cancellation of $U(1)$-nonabelian
or $U(1)$-gravitational mixed anomalies \cite{iru} (see \cite{abdss} 
for further discussions) do not impose further constraints beyond
(\ref{essential}). 

The remaining twisted RR tadpole cancellation conditions can however be
recovered by requiring cancellation of anomalies on fat strings obtained
by wrapping D3-branes (denoted D3$_p$) on e.g. the third complex plane,
with trivial Wilson lines, and located at the origin in the first two
planes. Such branes are fixed under the $\IZ_3$ action, and the general
form of the Chan-Paton action is
\beqa
\gamma_{\theta,3_p} & = & -\diag (\id_{n_{00}}, \id_{n_{01}},
\id_{n_{0,2}},
\alpha \id_{n_{10}}, \alpha \id_{n_{11}}, \alpha \id_{n_{12}}, \alpha^2
\id_{n_{20}}, \alpha^2 \id_{n_{21}}, \alpha^2 \id_{n_{22}}) \nonumber \\
\gamma_{\omega,3_p} & = & -\diag (\id_{n_{00}}, \alpha \id_{n_{01}},
\alpha^2 \id_{n_{0,2}}, \id_{n_{10}}, \alpha \id_{n_{11}}, \alpha^2
\id_{n_{12}}, \id_{n_{20}}, \alpha \id_{n_{21}}, \alpha^2 \id_{n_{22}})
\eeqa
The orientifold projection requires $n_{ab}=n_{-a,-b}$. We have
$\gamma_{\Omega',3_p}=-\gamma_{\Omega',3_p}^T$.

The field theory on the two-dimensional world-volume of the fat string is
$(0,2)$ supersymmetric. It contains a set of gauge multiplets (formed by
gauge bosons and two left-handed Majorana-Weyl fermions), chiral multiplets 
(containing one complex scalar and two right-handed MW fermions), and
Fermi multiplets (containing two left-handed MW fermions) (see
\cite{twodim,gcu} for a review of supermultiplet structure). 
In the $3_p3_p$ sector, we obtain the following set of multiplets
\beqa
\begin{array}{cc}
{\bf (0,2)}\; {\rm\bf multiplet} & {\rm \bf Representation} \\
{\rm Vector} & USp(n_{00}) \times U(n_{10}) \times U(n_{01})
\times U(n_{11}) \times U(n_{12}) \\
{\rm Chiral} & \Yasymm_{00} + {\rm Adj}_{10} + {\rm Adj}_{01} + {\rm
Adj}_{11} + {\rm Adj}_{12} \\
{\rm Chiral}  & \Ysymm_{10} + (\fund_{00},\antifund_{10}) +
(\fund_{01},\antifund_{11}) + (\fund_{11},\fund_{12}) +
(\antifund_{12},\antifund_{01}) \\
{\rm Fermi} & \Yasymm_{10} + (\fund_{00},\antifund_{10}) +
(\fund_{01},\antifund_{11}) + (\fund_{11},\fund_{12}) +
(\antifund_{12},\antifund_{01}) \\
{\rm Chiral} & \Ysymm_{01} + (\fund_{00},\antifund_{01}) +
(\fund_{10},\antifund_{11}) + (\fund_{11},\antifund_{12}) +
(\fund_{12},\antifund_{10}) \\
{\rm Fermi} & \Yasymm_{01} + (\fund_{00},\antifund_{01}) +
(\fund_{10},\antifund_{11}) + (\fund_{11},\antifund_{12}) +
(\fund_{12},\antifund_{10}) \\
{\rm Chiral} & \bYasymm_{11} + (\fund_{00},\fund_{11}) +
(\fund_{10},\fund_{01}) + (\antifund_{01},\fund_{12}) +
(\antifund_{12},\antifund_{10}) \\
{\rm Fermi} & 
\bYsymm_{11} + (\fund_{00},\fund_{11}) + (\fund_{10},\fund_{01}) +
(\antifund_{01},\fund_{12}) + (\antifund_{12},\antifund_{10}) 
\end{array}
\eeqa
In the $33_p+3_p3$ secto we have
\beqa
\begin{array}{cc}
{\bf (0,2)}\; {\rm\bf multiplet} & {\rm \bf Representation} \\
{\rm Chiral} & (\fund_{0,0};N_{11}) + (\fund_{01};N_{10}) +
(\antifund_{01};N_{12}) + (\fund_{10};N_{01}) + (\fund_{11};N_{00}) \\
 & (\antifund_{11};{\ov N}_{11}) + (\antifund_{12};{\ov N}_{10}) +
(\antifund_{10};{\ov N}_{12}) + (\fund_{12},{\ov N}_{01}) \\
{\rm Fermi} & (\fund_{0,0};N_{12}) + (\fund_{01};N_{11}) +
(\antifund_{01};N_{10}) + (\fund_{10};{\ov N}_{01}) + (\fund_{11};N_{01})
\\
& (\antifund_{11};{\ov N}_{10}) + (\antifund_{12};{\ov N}_{12}) +
(\antifund_{10};{\ov N}_{11}) + (\fund_{12},N_{00})
\end{array}
\eeqa
The field theory is chiral, and contains potential non-abelian gauge
anomalies. The conditions for their cancellation are
\beqa
N_{11}-N_{12} -4 & = & 0 \nonumber \\
N_{00} + N_{11} - N_{01} - N_{10} - 4 & = & 0
\label{probefour}
\eeqa
They amount to the conditions
\beqa
\Tr \gamma_{\theta\omega}= -4 \quad ; \quad \Tr \gamma_{\theta\omega^2}=8
\eeqa
Namely, besides (\ref{essential}), which was already required for
cancellation of four-dimenasional anomalies, consistency of the probe
requires the cancellation of the twisted tadpole associated to the third
fixed plane. Notice that the probe we have introduce is not sensitive to
the $\theta$- or $\omega$-twisted tadpoles, since they are not localized
at the origin in the first two complex planes (where our probe sits).
Clearly, they are required for consistency of other probes, obtained as
fat strings from D3-branes wrapped in one of these planes (and at the
origin in the other two).

The same results could have been derived by considering other probes, like
a D7-brane completely wrapped in $\IT^6$. Instead of pursuing their
study, we conclude our discussion of toroidal orientifolds by restating
our basic point. From the viewpoint of the compactified theory, the strong
constraints imposed by RR tadpole cancellation arise because they must
ensure not just the consistency of the low-energy field theory in the
vacuum, but also in non-trivial topological sectors.

\section{Application: Orientifolds of curved K3 manifolds}

\subsection{Smooth fibers}

The strategy of testing the consistent cancellation of RR charges in open
string vacua by studying probes sensitive to the relevant RR tadpoles can
be used in contexts where RR tadpole conditions cannot be directly computed 
using the familiar CFT rules (factorization of one-loop amplitudes). In
this Section we construct some simple examples of orientifolds of curved
manifolds, namely K3 manifolds, for which no exact CFT description is
available, and use the probe criterion to determine the configuration of
D-branes required to achieve RR tadpole cancellation.

In order to keep the models simple, we consider the K3 to be an elliptic
fibration over $\IP_1$, with a section. Let $[C]$ and $[f]$ denote the
homology classes of base and fiber, respectively. Let us consider modding
out type IIB theory on such K3 by the action $\Omega'=\Omega R
(-1)^{F_L}$, where $R$ acts as $R:[z,w]\to [-z,w]$ on the projective
coordinates of the base $\IP_1$, and leaves the elliptic fiber invariant.
This action preserves half of the supersymmetries, and has two fixed
points on the base, $z/w=0$ and $z/w=\infty$. We choose a generic K3 so
that no singular fiber sits at these points in the base, hence the 24
singular fibers group in 12 $\Omega'$-invariant pairs.

The closed string sector contains the $D=6$, $\NN=1$ supergravity and
dilaton tensor multiplet. In addition, out the 20 $(1,1)$ two-forms of K3
\footnote{Useful information on the structure of 2-cycles in elliptic K3's
can be extracted from \cite{aspinwall}.}, the base and fiber are invariant
and give two hypermultiplets. Out of the remaining 18, eight are
associated to paths in the base between locations in a set of eight
singular fibers, while eight are the 2-cycles associated to the $\Omega'$
image singular fibers. These cycles are exchanged by $\Omega'$, and
contribute eight hyper- and eight tensor multiplets. Finally, two 2-cycles
correspond to paths between a singular fiber and its image. These cycles
are invariant and contribute two hypermultiplets.

Clearly, consistency requires the introduction of a number of D7-branes
wrapped on the fiber and sitting at points on the base. We locate them
away from the fixed points on the base and from singular fibers. Since
their tangent and normal bundles are trivial ($f$ is a two-torus, 
and $[f]\cdot [f]=0$), the D7-branes are quite insensitive to the
curvature of K3, and lead to $D=6$, $\NN=2$ supersymmetric non-chiral
spectra. Denoting by $N$ the number of $\Omega'$-pairs of D7-branes,
at a generic point in moduli space the open string sector contributes with
$U(1)^N$ vector multiplets, and $N$ hypermultiplets. The full (open plus
closed) spectrum is free of six-dimensional gauge and gravitational
anomalies, regardless of the  number of D7-branes in the model. 

This number is fixed by cancellation of RR charges in the compact space.
Instead of determining it directly, we may obtain it by demanding consistency 
of the world-volume theory on a suitable probe, taken to be a fat string
obtained by wrapping $n$ D3-branes on the base $\IP_1$, the section of the
fibration. In the absence of curvature the two-dimensional probe
world-volume would have $(0,8)$ supersymmetry, but the K3 holonomy reduces
it to $(0,4)$. In the $33$ sector we obtain $(0,4)$ $SO(n)$ vector multiplets 
(containing one gauge boson, and four left-handed MW spinors), and one
$(0,4)$ chiral multiplet (containing two complex scalars and four
right-handed MW spinors) in the two-index symmetric representation. The
latter parametrizes motion of the fat string in four transverse spacetime
dimensions. Note that we do not obtain chiral multiplets associated to
internal dimensions in K3 because the curve $C$ wrapped by the D3-branes
is rigid, $[C]\cdot [C]=-2$, forbidding transverse motion, and has no
constant holomorphic 1-forms, forbidding Wilson lines. In the 37+73
sectors, we find, for each of the $N$ $\Omega'$-invariant pairs of
D7-branes, two left handed MW spinor in the fundamental of $SO(n)$.
Cancellation of two-dimensional anomalies then requires
\beqa
-4 \times \frac {n-2}{2} + 4\times \frac {n+2}{2} - 2\times N\times \frac
12 = 0
\eeqa
fixing $N=8$. Hence the model requires 16 D7-branes (as counted in the
covering space) for consistency.

In this simple example, the final result could be obtained by noticing
there are two O7-planes in the model, which wrap flat curves with trivial
normal bundles, and therefore carry the same charges as in flat space,
namely $-8$ units of D7-brane charge for each. To cancel their
contribution 16 D7-branes must be introduced, as found above.

Incidentally, the geometry of D7-branes and O7-planes is so simple that it 
is possible to construct the F-theory \cite{vafafth} lift of this model.
Since D7-branes and O7-planes wrap the flat elliptic fiber $f$, along
which there are two $U(1)$ isometries, these directions can be ignored and
the problem reduces to lifting two O7-planes and eight D7-branes sitting
on $\IP_1/\IZ_2$, which is another $\IP_1$. The F-theory lift produces
another elliptic fibration over $\IP_1$, with fiber denoted by $f'$. The
number of singular fibers is 12, eight arising from the D7-brane pairs in
the orientifold quotient, and four arising from the splitting of the each
O7-plane into two mutually non-local sevenbranes \cite{senfth}. The final
result is an F-theory compactification on a threefold, obtained as a double 
elliptic fibration over $\IP_1$. As just described, the F-theory fibration
with fiber $f'$ has 12 degenerate fibers, and so has the original fibration, 
with fiber $f$ (arising from the original 24 singular fibers in K3, identified 
in pairs by the orientifold action). These are precisely the correct numbers 
to render the threefold Calabi-Yau and preserve eight supercharges. This
threefold has appeared in \cite{vafafth}, and the generic spectrum of the
model ($\NN=1$ supergravity, nine tensor multiplets, eight vector multiplets 
and twenty hypermultiplets) agrees with that obtained above. The correct
F-theory lift serves as a further check of the consistency of the model,
and the validity of the probe criterion in general situations.

\subsection{Degenerate fibers}

The above model can be modified to yield more interesting six-dimensional
physics, with chiral gauge sectors. Also, the models below present
additional subtleties in the computation of orientifold-plane charges,
hence illustrating the usefulness of the probe approach in certain
situations.

Let us consider the same $\Omega'$ orientifold of an elliptic K3, but in
which the fiber over one of the fixed points of $\Omega'$ in $\IP_1$, say
$z=\infty$, is smooth, and the fiber over $z=0$ is singular, necessarily
at least of type $I_2$ in Kodaira classification (see \cite{mv}). Namely,
the two-torus is pinched twice so that topologically it is a set of two 
two-spheres $C_1$ and $C_2$ touching at two points. The closed string 
spectrum differs from the generic one in yielding one less hypermultiplet
and one more tensor multiplet. The closed string sector leads to gravitational 
anomalies, which must be cancelled by the open string sector. 

Since the orientifold plane is wrapping a reducible curve it may possess
different charges under the RR forms obtained by integrating over $C_1$
and $C_2$. This is confirmed by considering the local configuration of an
O7-plane wrapped on an $I_2$-degenerated elliptic fiber in K3 and T-dualizing 
along the unpinched $U(1)$ orbit as in \cite{tdual}. One obtains a 
configuration of two NS fivebranes spanning the directions 012345 and one 
O6-plane along 0123456, in a locally flat space with the direction 6
compactified on a circle. These configurations have been studied in 
\cite{bkhz} (see \cite{lll} for a four-dimensional version) in the context
of Hanany-Witten brane constructions \cite{hw}. In this situation the 
O6-plane is known to flip charge as it crosses the NS-branes in the $x^6$
direction \cite{ejs}. Hence the original O7-plane carries opposite charges
in the two components $C_1$, $C_2$ in the $I_2$ fiber \footnote{The 
configuration seems to be related, in the limit of one collapsing cycle,
to an orientifold of the $\IC^2/\IZ_2$ singularity studied in \cite{upermu}. 
This picture allows a direct computation of these charges, open string
spectrum, and twisted tadpoles below.}.

We may expect that the charge difference in the two components must be
compensated by D7-branes wrapped on them, in different numbers for $C_1$
and $C_2$. In fact, this nicely follows from cancellation of six-dimensional 
gauge anomalies on the corresponding D7-brane world-volume. Let us
consider $N_1$, $N_2$ D7-branes wrapped on $C_1$, $C_2$, and sitting at
$z=0$. Using the T-dual picture mentioned above, the open string sector
leads to the following $D=6$ $\NN=1$ multiplets
\beqa
{\rm Vector} & SO(N_1)\times USp(N_2) \nonumber \\
{\rm Hyper} & (N_1,N_2)
\eeqa
Cancellation of six-dimensional anomalies implies $N_1-N_2=8$, but does
not constrain the overall number of D7-branes. This condition also ensures
cancellation of gravitational anomalies. Notice the analogy of the
condition with twisted RR tadpole conditions of the type encountered in
Section 3.1 (in fact the condition arises as a twisted tadpole conditions
in the $\IC^2/\IZ_2$ orientifold mentioned in the footnote 6).

There remains to fix the total number of D7-branes in the model. Notice
that a direct computation of the orientifold charge in this case is rather
subtle, and that the F-theory lift seems rather involved. Happily, the
number 
$N$ of D7-branes pairs away from $z=0$ can be obtained by considering the
same D3-brane probe as before. Assume for simplicity that the section $C$
intersects the component $C_1$, rather than $C_2$. Then the 33 spectrum is
as above, and the number of chiral fermions in the fundamental representation 
in $37+73$ sectors is $N_1+2N$, so we obtain
\beqa
N_1 + 2N = 16
\eeqa
Actually, the result can be confirmed by considering a transition in which
$C_2$ is shrunk to zero size and the $I_2$ degenerate fiber can be sent to
non-zero $z$ as two $I_1$ fibers related by $\Omega'$. We are left with
$N_1$ D7-branes wrapped on the smooth fiber over $z=0$, which can also be
sent to $z\neq 0$ as $N_1/2$ $\Omega'$ invariant pairs. The final model is
of the type studied in section 4.1, with $N+N_1/2$ D7-brane pairs, hence
consistency requires $2\times (N+N_1/2)=16$, as found here. The process
just described is T-dual to that studied in \cite{bkhz}. 

\smallskip

The model admits a simple generalization, by considering $I_{2n_1}$, 
$I_{2n_2}$  degenerations of the fibers over $z=0$, $z=\infty$. The closed
string spectrum contains the supergravity and dilaton tensor multiplet, 
and $12-n_1-n_2$ hyper- and $8+n_1+n_2$ tensor multiplets. Cancellation of
tadpoles associated to cycles in the reducible fibers requires to locate at 
least 8 D7-branes wrapped on suitable components in the fibers over $z=0$,
$z=\infty$. The number of D7-pairs away from the orientifold points can be
seen to be zero, by demanding consistency of a D3-brane probe. The
resulting open string spectrum is simply given by $SO(8)^{n_1+n_2}$ vector
multiplets, and the full spectrum is free of gauge and gravitational
anomalies. Some of these spectra coincide with those in \cite{blum}, and
it would be interesting to explore for more concrete connections with
them.

Another interesting extension would be to consider other degenerated
fibers over the orientifold fixed points, which presumably lead to gauge
sectors of the type constructed in \cite{bi2,upermu}. We leave these and
other possibilities for further research.

\section{K-theory charges and D-brane probes}

Traditionally the consistency conditions imposed on theories with open
string sectors has been cancellation of RR charges in a compact space,
understood as a condition in (co)homology. As explained in the introduction, 
the spacetime argument leading to this constraint is consistency of the
equation of motion for the RR $p+1$-form field in the presence of D-brane
sources $\delta(W_i)$ 
\beqa
d*H_{p+1} = \sum_i \delta(W_i)
\label{kth}
\eeqa
when taken in cohomology. Recent developments on non-supersymmetric states
in string theory (see \cite{revnbps} for a review) have shown that D-brane
charges are however not classified by (co)homology, but by K-theory
\cite{witten}. There arises the question of whether consistency of a
compactification requires cancellation of D-brane K-theory charge or
merely its cancellation in cohomology. 

To give a concrete example, there exist certain D-brane states in type I
string theory which do not carry standard RR charges, but have $\IZ_2$
K-theory charges \cite{senspinor,witten}. Since there is no field carrying
these charges, the above argument does not seem to apply, and
compactifications with an odd number of objects of this kind would seem
consistent. On the other hand, given a compact space ${\bf X}$ and any
topological  charge sitting a point $P$ in it, we may split ${\bf X}$ into
two pieces ${\bf X}_1$ and ${\bf X}_2$, both with boundary ${\bf Y}$ (but
with opposite orientations) and with $P\in {\bf X}_1$. Physically, it must
be possible to measure the topological charge in ${\bf X}_1$ by measuring
quantities in ${\bf Y}$, and since ${\bf Y}$ is also the boundary of
${\bf X}_2$, with opposite orientation, it follows that the opposite
charge is contained in ${\bf X}_2$. Consistency must require cancellation
of full D-brane charges, i.e. K-theory charges.

In fact the criterion of measuring K-theory charges by looking at
asymptotics of the configuration has lead in \cite{mw} to the conclusion
that RR fields are also described by K-theory classes and hence (\ref{kth}) 
applies in K-theory. This description and a detailed analysis of Dirac
quantization conditions provided a beautiful unified explanation of the
different shifted flux quantization conditions for RR fields.

The conclusion is however exotic from the point of view of usual construction 
of orientifold models, where the familiar procedure to evaluate RR tadpole
consistency conditions by factorizing one-loop amplitudes seems sensitive
only to the cohomology part of D-brane charge. We leave the question of
analyzing possible subtleties in this procedure in the presence of K-theory 
charges beyond (co)homological ones as an open issue. Instead, in this
Section we present an alternative point of view on the discussion of
cancellation of K-theory charge, with arguments more concrete than the
above rather formal ones. In particular we consider some simple
compactifications where standard RR charge is cancelled but K-theory 
charge is not, and show that the corresponding inconsistency shows up as
global gauge anomalies \cite{global} in suitable D-brane probes.

\subsection{Toroidal compactification of type I with non-BPS D7-branes}

Let us consider the simplest case. We consider a compactification of type
I theory on $\IT^2$, with a single non-BPS D7-brane (denoted \Dtseven-brane) 
spanning the eight non-compact dimensions and sitting at a point in $\IT^2$. 
The type I \Dtseven-brane does not carry standard RR charges, but carries
a K-theory $\IZ_2$ charge \cite{witten}. It also contains a tachyonic
mode, arising from the 79, 97 open strings \cite{lerda}, but this fact
does not affect the topological properties of the configuration, and we
safely ignore it. 

Notice that since the \Dtseven-brane is not charged under the RR fields
the inconsistency of the configuration is not too obvious. However, it can
be shown as follows. Recall that the $\IZ_2$ \Dtseven-brane charge can be
detected by carrying a non-BPS D0-brane around a small circle surrounding
the D7-brane location in $\IT^2$, the D0-brane wavefunction changes sign
as proposed in \cite{witten} and computed in \cite{sergei}. But if no
other $\IZ_2$ charge sits on $\IT^2$ the contour can be deformed to a
small circle around the `other side' of $\IT^2$, leading to a sign flip in
the D0-brane wavefunction without surrounding any $\IZ_2$ source, hence an
inconsistency. This proves that \Dtseven-branes in compact spaces must
exist in pairs so that their K-theory charge cancels \footnote{One can
operate similarly for
other K-theory charges. For instance, the non-BPS D(-1)-brane amplitudes
are weighted by opposite signs on the two sides of a non-BPS type I
D8-brane \cite{witten,sergei}. This implies that e.g. $\IT^2$ type I
compactifications with a single D8-brane are inconsistent due to the
impossibility to define `sides' consistently. The number of D8-branes must
be even, i.e. the corresponding $\IZ_2$ charge cancels in the
compactification.}

Let us offer another argument based on the use of D-brane probes similar
to those in previous sections. Consider probing the previous configuration
by a set of $2n$ coincident D5-branes wrapped on $\IT^2$. Let us compute
the fields propagating on the four non-compact dimensions. The 55, 59 and
95 sectors preserve $\NN=2$ supersymmetry, and lead to the following
multiplets (as in \cite{witinst})
\beqa
\begin{array}{ccc}
{\bf 55} & \NN=2\; {\rm Vector} & USp(2n) \\
         & \NN=2\; {\rm Hyper}  & \Yasymm \\
{\bf 59+95} & \NN=2\; {\rm Hyper} & \frac 12 (2n;32)
\end{array}
\eeqa
the latter decomposed in suitable irreducible representations if $SO(32)$
Wilson lines are turned on. The interesting piece arises from the $5\hat
7$ and $\hat 7 5$ sectors, which are non-supersymmetric. Since the
\Dtseven-brane is constructed as a type IIB D7-\Dseven pair exchanged by
$\Omega$ \cite{witten},
the $57$ and $5\bar 7$ sectors map to the $\bar 7 5$ and $75$ sector, so
we simply compute the spectrum in the former and do not impose the $\Omega$ 
projection. Centering on the fermion content, we find one Weyl fermion in
the fundamental representation $2n$ of $USp(2n)$. 

Even though the resulting field theory is non-chiral, it is inconsistent
at the quantum level due to a global gauge anomaly \cite{global}. For
compactifications with an arbitrary number $N$ of \Dtseven-branes, the
problem is absent precisely when $N$ is even, namely when the \Dtseven-brane 
$\IZ_2$ K-theory charge cancels. Hence, cancellation of torsion pieces of
K-theory charge can be detected by global gauge anomalies on suitable
D-brane probes. We hope this argument makes K-theory considerations more
familiar and tractable in the construction of orientifolds, a desirable
aim in view of recent introduction of non-BPS branes in type IIB
orientifold models \cite{abg,ru}.

\subsection{K-theory charge in type IIB $\IT^4/\IZ_2$ orientifolds}

In this section we would like to present a more involved but very interesting 
(and related) example. It is based on the $\Omega$ orientifold of type IIB
theory on $\IT^4/\IZ_2$ constructed in \cite{bisagn} and \cite{gp}. The
closed string sector gives the $D=6$ $\NN=1$ supergravity multiplet, the
dilaton tensor multiplet, and 20 hypermultiplets. The model contains 32
D9-branes, which for concreteness we consider without Wilson lines, and a
total of 32 D5-branes, which can be located in groups of $k_i$ pairs,
$i=1,\ldots,16$, $\sum_{i=1}^{16} k_i=16$, at the sixteen $\IZ_2$ fixed points
$(z_1,z_2)$ with $z_i=0,1/2,i/2, (1+i)/2$. The Chan-Paton embeddings can
be taken 
\beqa
\begin{array}{ll}
\gamma_{\theta,9} = \diag(i\id_{16},-i\id_{16}) &
\gamma_{\theta,5,i} = \diag(i\id_{k_i},-i\id_{k_i}) \\
\gamma_{\Omega,9} = {\pmatrix{ & \id_{16} \cr \id_{16} & \cr}} &
\gamma_{\Omega,5,i}  =  {\pmatrix{ & \id_{k_i} \cr -\id_{k_i} & \cr}}
\end{array}
\eeqa
The open string spectrum of the model has the following $D=6$ $\NN=1$
multiplets
\beqa
\begin{array}{ccc}
{\bf 99} & {\rm Vector} & U(16) \\
         & {\rm Hyper}  & 2\; \Yasymm \\
{\bf 55} & {\rm Vector} & \prod_{i=1}^{16} U(k_i) \\
         & {\rm Hyper}  & \sum_{i=1}^{16} 2\;\Yasymm_{\,i} \\
{\bf 59+95} & {\rm Hyper} & \sum_{i=1}^{16} (\fund_{\,i};16)
\end{array}
\eeqa

Any choice of $k_i$'s with $\sum_{i=1}^{16} k_i=16$ cancels the RR tadpoles 
and would seemingly lead to a consistent model. However, it was observed in 
\cite{blpssw} that there is a further constraint on the possible 
distributions. Careful analysis of Dirac quantization conditions leads to
the constraint that the total number of D5-brane pairs at any four $\IZ_2$
fixed points lying on a two-plane  must be a multiple of two. That is 
$\sum_{i'=1}^4 k_{i'}=0$ mod $2$, with $i'$ labeling fixed points in the
plane. In a T-dual picture, the conditions could be rephased as the conditions 
that Wilson lines on the dual D9-branes should allow the existence of
spinor representations, which were known to exist in the model due to type
I/heterotic duality. Since at the time the type I description of these states 
was not known, the conditions were suggested to be of non-perturbative
origin. With our present knowledge \cite{senspinor}, states in the spinor
representation arise from type I non-BPS D0-branes, and the conditions
above correspond to requiring a well-defined wavefunction for such states.
Namely, it corresponds to cancellation of K-theory \Dtseven-brane charge
(which appears induced on the D9-branes by the Wilson lines). Hence the 
requirements in \cite{blpssw} are K-theory charge cancellation conditions,
stated before the advent of K-theory.

We conclude by describing how the inconsistency arises in a suitable
D-brane probe in the original picture. In this case the use of the D-brane
probe argument is easier than the original argument in \cite{blpssw}, and
we expect it to be more general and useful in more complicated models, or
models without geometrical interpretation for the internal space.

Let us introduce a D5-brane, denoted D5$_p$-brane to distinguish it from
those in the background, wrapped on a two-plane passing through four
$\IZ_2$ fixed points, labeled by $i'$, in $\IT^4/\IZ_2$. Consistency of
the $\IZ_2$ projection in mixed $5_p5$, $55_p$, $5_p9$, $95_p$ sectors
requires $\gamma_{\theta,5_p}^2=\id$. The Chan-Paton embeddings have the
general form $\gamma_{\theta,5_p}=\diag(\id_{n_0},-\id_{n_1})$ and
$\gamma_{\Omega,5_p}=\diag(\varepsilon_{n_0},\varepsilon_{n_1})$. The
probe preserves half of the supersymmetries and has $\NN=1$ susy in its
four non-compact world-volume dimensions. We obtain the following set of
multiplets
\beqa
\begin{array}{ccc}
{\bf 5_p5_p} & {\rm Vector} & USp(n_0)\times  USp(n_1) \\
             & {\rm Chiral} & \Ysymm_0 + \Ysymm_1 +
2\times(\fund_0,\fund_1)\\
{\bf 5_p9 + 95_p} & {\rm Chiral} & (\fund_0;16)+(\fund_1;16) \\
{\bf 5_p5 + 55_p} & {\rm Chiral} & (\fund_0;k_{i'})+
(\fund_1;k_{i'}) 
\end{array}
\eeqa
Cancellation of global gauge anomaly \cite{global} leads to the constraint
$\sum_{i'}k_{i'}=0$ mod $2$, the consistency condition mentioned above.

Hence, we see the additional consistency condition in \cite{blpssw} can
actually be detected by using by now familiar D-brane probes. Moreover, we
have seen in Section 5.1 that global anomalies on D5-brane probes wrapped
on two-tori are related to K-theory charge of D7-branes transverse to that
plane. Our probe analysis agrees with our previous interpretation of
the additional consistency conditions as K-theory charge cancellation
conditions. We hope this technique is helpful in studying such consistency
conditions in other type IIB orientifolds, or in more general string vacua
with open string sectors.

\section{Conclusions}

In this paper we have considered the introduction of D-brane probes in
diverse string compactifications with open string sectors, and we have
studied the interplay between world-volume gauge anomalies and charge
cancellation consistency conditions. We have found that RR tadpole
cancellation conditions are equivalent to cancellation of chiral gauge
anomalies on suitable probes one may introduce. We find this viewpoint
interesting since it provides a rationale, from the compactified effective
theory viewpoint, of the strong constraints implied by tadpole cancellation, 
typically much stronger than cancellation of anomalies in the compactified
theory vacuum. A second nice feature of the probe criterion is its wide
applicability, which we have illustrated with the construction of
consistent orientifolds of curved K3 spaces.

We have employed D-brane probes to explore charge cancellation consistency
conditions in compactifications involving torsional K-theory charges. We
have found that uncancelled charges show up as global gauge anomalies on
suitable D-brane probes, and hence consistency requires cancellation of
full K-theory charges, in agreement with conclusions from formal
considerations \cite{mw}. Also, quite surprisingly, we have found non-trivial 
K-theory charge cancellation constraints in seemingly innocent type IIB
orientifolds, like the $\IT^4/\IZ_2$ model in \cite{bisagn,gp}.

For the purposes of this paper it has been enough to employ a relatively
small set of brane probes. In principle, extension to other kinds of
probes, either supersymmetric or non-supersymmetric (non-BPS branes,
stable or not, or brane-antibrane pairs, either spacetime-filling or not)
does not involve any conceptual difference. However, they may lead to
simpler analysis in certain situations, hence a more systematic
exploration would be desirable.

Along this line, we would like to point out that at present we do not have
a systematic characterization of the kind of probe sensitive to a specific
tadpole. It would be useful to provide a more detailed map between 
tadpoles and the corresponding probes. A possible guideline in geometric
examples seems to be provided by the intersection pairing in (co)homology
or K-theory. It would be interesting to make this idea more explicit in
singular geometries, such as those typically encountered in toroidal
$\IT^n/\IZ_N$ orientifolds, and to extend it to non-geometric 
compactifications (by using the CFT index introduced in \cite{df} or a
suitable generalization). We hope to address some of these issues in
future work.

\bigskip

\centerline{\bf Acknowledgements}

It is my pleasure to thank G.~Aldazabal, S.~Franco, L.~E.~Ib\'a\~nez,
J.~F.~Morales, Y.~Oz and R.~Rabad\'an for useful conversations. I also
thank the Universidad Aut\'onoma de Madrid for hospitality during the
completion of part of this work, and M.~Gonz\'alez for encouragement and
support.

\end{document}